\documentclass[prl,amsmath,amssymb,twocolumn]{revtex4-1}
\usepackage[normalem]{ulem}
\usepackage{amsmath,amssymb}
\usepackage[bookmarks=true,colorlinks,linkcolor=blue,urlcolor=blue,citecolor=blue]{hyperref}
\usepackage{graphicx,amsmath,amsfonts}
\usepackage[usenames]{color}
\usepackage{epstopdf}
\usepackage{verbatim}
\usepackage{amsthm}
\usepackage{enumitem}

\newcommand{\be}{\begin{equation}}
\newcommand{\beq}{\begin{eqnarray}}
\newcommand{\eeq}{\end{eqnarray}}

\def \tr{{\mbox{tr~}}}

\def \ua{{\uparrow}}
\def \da{{\downarrow}}
\def \be{\begin{equation}}
\def \ee{\end{equation}}
\def \ba{\begin{array}}
\def \ea{\end{array}}
\def \bea{\begin{eqnarray}}
\def \eea{\end{eqnarray}}

\def \half{\frac{1}{2}}

\def \W{{\Omega}}

\def \t{{\theta}}
\def \b{{\beta}}

\def \D{{\Delta}}
\def \d{{\delta}}

\def \G{{\Gamma}}

\def \nd{{^{\vphantom{\dagger}}}}
\def \yd{^\dagger}
\def \av#1{{\langle#1\rangle}}
\def \ket#1{{\,|\,#1\,\rangle\,}}
\def \bra#1{{\,\langle\,#1\,|\,}}

\def \eff{{\text{eff}}}

\begin{document}

%\title{Real-Space Renormalization Group for the dynamics of random spin chains and many-body localization}
\title{Many-body localization in one dimension as a dynamical renormalization group fixed point}
\author{Ronen Vosk and Ehud Altman \\{\small \em Department of Condensed Matter Physics, Weizmann Institute of Science, Rehovot 76100, Israel}}
\begin{abstract}
We formulate a dynamical real space renormalization group approach to describe the time evolution  of a random spin-$\half$ chain, or interacting fermions, initialized in a state with fixed particle positions. Within this approach we identify a many-body localized state of the chain as a dynamical infinite randomness fixed point. Near this fixed point our method becomes asymptotically exact, allowing analytic calculation of time dependent quantities. In particular we explain the striking universal features in the growth of the entanglement seen in recent numerical simulations: unbounded logarithmic growth delayed by a time inversely proportional to the interaction strength. The particle number fluctuations by contrast exhibit much slower growth as $\log\log t$ indicating blocked particle transport.
Lack of true thermalization in the long time limit is attributed to an infinite set of approximate integrals of motion revealed in the course of the RG flow, which become asymptotically exact conservation laws at the fixed point. Hence we identify the many-body localized state with an emergent generalized Gibbs ensemble.
\end{abstract}
\maketitle
%\section{Introduction}

%It was long believed that true insulators exist only at zero temperature. Even if particles are decoupled from lattice phonons the common expectation had been that collective excitations in an interacting system can play the same role and assist in energy an particle transport. Therefore the recent theoretical of a many-body localization transition of a quantum system at non vanishing temperature was one of the most intriguing predictions of recent years\cite{Basko}. The common wisdom had been that even if the particles are decoupled from lattice phonons or other external baths, collective excitations of an interacting system play essentially the same role and assist in particle and energy transport.

What is the effect of interactions on Anderson localization? One common wisdom is that any amount of interaction will give rise to collective excitations that could assist transport at non-vanishing temperature even if single particle states are all localized. But the belief, that there are no strict many-body insulators at $T>0$, has been challenged by theoretical arguments, dating as far back as Anderson's original paper, which suggest a many-body localization transition marking a critical point in the transport properties of a closed quantum system\cite{Anderson1958,Basko2006}. The idea has recently gained support from numerical studies \cite{Oganesyan2007,Monthus2010,Pal2010,Canovi2011}. Furthermore, simulations of one dimensional systems have revealed remarkably universal behavior of the dynamics in the putative many-body localized state\cite{Chiara2006,Znidaric2008,Bardarson2012}. For example the time evolution following a quench from a state with fixed particle positions shows blocked particle transport accompanied by unbounded logarithmic growth of the entanglement entropy.

In this paper we formulate a real space renormalization group (RG) scheme that describes the time evolution of a random spin chain on multiple time scales. For certain initial conditions we can establish a many-body localized state as an infinite randomness fixed point of the dynamics near which the RG scheme is asymptotically exact. The results of this theory exhibit many of the universal features observed in the numerical simulations\cite{Chiara2006,Znidaric2008,Bardarson2012}, although they are derived for a somewhat different model.
As our starting point we consider the Hamiltonian of the random spin-$1/2$ XXZ chain without local Zeeman fields:
\be
H = \sum_i \frac{J_i}{2} \left(S^+_i S^-_{i+1}+S^-_i S^+_{i+1}+2 \D_i S^z_i S^z_{i+1}\right).
\label{eq:Hamiltonian}
\ee
The couplings $J_i$ and anisotropy parameters $\D_i$ on sites $i$ are random variables drawn from uncorrelated probability distributions. The couplings may be positive or negative and we assume $|\Delta_i|<1$. The spins in (\ref{eq:Hamiltonian}) can be mapped using a Jordan-Wigner transformation to spinless fermions with nearest neighbor interactions, subject to bond disorder.

To study the propagation of information through the chain we investigate the time evolution of the system starting from a non entangled initial state, which for simplicity we take as an  antiferromagnetic Ne\'el state with spins pointing along the $z$-axis. We shall see that this choice of initial state greatly simplifies the scheme and allows us to obtain well controlled results for the dynamics at long times.

\noindent{\em RG scheme --} Before proceeding with the details, let us outline the idea underlying the RG solution of the time evolution. As in the standard strong disorder RG scheme\cite{Dasgupta1980,Bhatt1982,Fisher1994}, we take advantage of the large local separation of energy scales induced by the randomness to gradually eliminate degrees of freedom. However, instead of targeting the ground state of the system, our aim is the time evolution of the chain starting from a specified initial state. The dynamics at the shortest time scales are oscillations of frequency $\W$ performed by the most strongly coupled pairs of spins on the chain, which are effectively decoupled on these time scales from their typically much slower neighbors. On time scales longer than $\W^{-1}$ the rapid oscillations performed by the strong bonds can be eliminated using time dependent perturbation theory, which leads to renormalization of the coupling between the slow spins. In this way we gain the essential information about the dynamics of the chain at all scales. If the distribution of coupling constants flows to a wide distribution at the dynamical fixed point, then the perturbative RG approach becomes increasingly well controlled, or even asymptotically exact if the system flows to infinite randomness\cite{Fisher1994}. Similar ideas have been applied to solve classical dynamics in certain disordered\cite{Fisher1998,Lee2009} as well as clean\cite{Mathey2010} systems, but to our knowledge not to quantum dynamics.

We now apply this scheme to the Model (\ref{eq:Hamiltonian}) with a staggered (Ne\'el) initial state. The strong bond, with exchange coupling denoted by $\W$, then always connects a pair of anti-aligned spins. At time scales of order $\Omega^{-1}$ the strong pair is effectively decoupled from the neighboring spins that are presumably connected to it by much weaker couplings. The only significant dynamics in this neighborhood of the chain is the rapid oscillation of the strong pair with frequency $\W$ between the $\ket{\ua\da}$ and $\ket{\da\ua}$ states.
%where  $\ket{\pm}=(\ket{\ua\da}\pm\ket{\da\ua})/\sqrt{2}$

The effective Hamiltonian for the dynamics at time scales much larger than $\W^{-1}$
is derived using time dependent perturbation theory in the coupling of the strong pair to the rest of the chain. Specifically, we move to the interaction picture with respect to the Hamiltonian of the strong pair and compute the evolution of the density matrix $\rho(t)= U_I\yd(t)\ket{\psi_0}\bra{\psi_0} U_I\nd(t)$ up to second order, while averaging over rapid oscillations of frequency $\W$. The resulting time evolution can be matched term by term to an effective time evolution $\exp(-iH_{\text{eff}}t)\ket{\psi_0}\bra{\psi_0} \exp(iH_{\text{eff}}t)$. Therefore this procedure amounts to derivation of an effective Hamiltonian.

At this order of perturbation theory the $\ket{\ua\ua}$ and $\ket{\da\da}$ are not populated and therefore truncated from the Hilbert space. The retained states $\ket{\pm}=2^{-1/2}(\ket{\ua\da}\pm \ket{\da\ua})$ of the strong pair can be taken as the $\ua/\da$ states of a new pseudo spin variable $\vec{S}_n$, which initially points along the positive or negative $x$-axis. The effective Hamiltonian derived in this procedure for our model is then given by:
\be \label{eq:Heff}
\begin{split}
H_{\text{eff}} &=H_\text{chain} + h_n S^z_n+\frac{J_L J_R}{2\Omega(1-\D_S^2)}\left(S_L^{+} S_R^{-} + S_L^{-} S_R^{+}\right) \\
&+\frac{\D_S J_L J_R}{2\Omega } \left[\frac{S_L^{+} S_R^{-} + S_L^{-} S_R^{+}}{(1-\D_S^2)}-\frac{\D_L \D_R}{\D_S}S_L^z S_R^z\right]S_{n}^z
\end{split}
\ee
Because $[H_\eff,S^z_n]=0$ the time evolution can be computed separately for each eigenvalue $\pm\half$ of $S^z_n$, using $H_{\text{eff}}^{\pm}$ that does not depend on the operator $\vec{S}_n$. The different evolution under $H_\eff^\pm$ together with the fact that the  new spin starts in a superposition of $\ua$ and $\da$ leads to entanglement between the effective-spin on the strong bond and its two neighboring spins. Full entanglement is generated after a time $t_{\text{ent}} =  2\Omega/(J_L J_R \D_S)$, set by the difference in the exchange constant in $H_\eff^\pm$.
This process will be important later on for computing the evolution of the entanglement entropy.

Apart from generating entanglement, the difference between the evolution given $\ua_n$ or $\da_n$ is not  crucial for the subsequent dynamics in the sense that they both lead to the same recipe for
renormalization of coupling constants. $H_{\text{eff}}^+$ and $H_{\text{eff}}^-$ have the same form as the original hamiltonian and we can directly read off the coupling
generated between $\vec{S}_L$ and $\vec{S}_R$, neighboring the strong bond to the left and right, upon decimation of that bond:
$\tilde{J} ={J_L J_R}/{\Omega}$ and $\tilde{|\D|} = {|\D_L| |\D_R|}/{4}$,
where we neglected the linear $\D$ correction to $\tilde{J}$. This approximation will be justified a posteriori by the fact that $\D$ flows to zero. The renormalization of the exchange coupling is then identical to that found in the random Heisenberg chain at $T=0$ and leads to the Random singlet phase\cite{Dasgupta1980,Bhatt1982,Fisher1994}. Note also that we keep only the absolute value of the anisotropy. The sign will randomize in the course of the RG flow because it depends on the state of $S^z_n$. These RG steps are now iterated as we gradually approach lower frequency scales and longer times.

The RG steps are iterated to produce a flow of the probability distributions with decreasing frequency cutoff $\W$ starting from the microscopic cutoff $\W_0$. Using the scaling variables
$\zeta = \ln \frac{\Omega}{J}$ and $\b = -\ln |\Delta|$,
and $\Gamma = \ln({\Omega_0}/{\Omega})=\ln(\W_0 t)$ we obtain the following equation for the joint probability distribution $P(\zeta,\beta;\G)$:
\begin{widetext}
\be \label{eq:general-flow}
\begin{split}
\frac{\partial P}{\partial \Gamma} &= \frac{\partial P}{\partial \zeta}
  + \rho(0;\G) \int_0^\infty d\b_L  d\b_R d\zeta_L d\zeta_R \delta(\zeta-\zeta_L-\zeta_R)\delta(\b-\b_L-\b_R-\ln4)P(\zeta_L, \b_L;\G)P(\zeta_R, \b_R;\G).
\end{split}
\ee
\end{widetext}
where $\rho(\zeta;\G)=\int d\b P(\zeta, \b;\G)$ is the distribution of $\zeta$.
Note that even if initially the variables $\zeta$ and $\b$ are independent a correlation builds up in the course of renormalization in the same way as it is generated in the ground state \cite{Fisher1994}.

By integrating over $\b$ we obtain an equation for $\rho(\zeta;\G)$,
\be
\frac{\partial \rho}{\partial \Gamma} = \frac{\partial \rho}{\partial \zeta} +\rho(0;\G) \int_0^\infty d\zeta_L d\zeta_R \delta(\zeta-\zeta_L-\zeta_R) \rho(\zeta_L) \rho(\zeta_R).
\ee
This equation is identical to the flow leading to the random-singlet ground state and it is solved by the same ansatz\cite{Fisher1994}, $\rho(\zeta;\G) = a(\Gamma) e^{-a(\Gamma) \zeta}$ with
\be \label{eq:solution}
\begin{split}
a(\Gamma) &= \frac{1}{\Gamma + 1/a_0}.
\end{split}
\ee
%which describes a flow to an infinite randomness fixed point.

%The RG fixed point is fully described by a fixed point joint distribution $P_\Gamma(\zeta, \b)$. The existence and asymptotic behavior of the fixed point distribution was obtained in Ref. \onlinecite{Fisher1994}, showing that typical $\b$ grows like $\Gamma^\phi$, where $\phi=(1+\sqrt{5})/2$ is the golden ratio. Hence, the flow of the typical value of $\D$ is toward zero, which justifies our previous approximations taking $\Delta\ll 1$.

Of course the above solution includes only partial information on the fixed point of the dynamics. Important information for calculation of physical quantities is held in the conditional average of the interaction variable $\b$ given a value of $\zeta$ on the same bond, ${\bar\b}(\zeta,\Gamma) \equiv \int_0^\infty d\b \b P (\zeta, \beta;\G)/\rho(\zeta;\G)$. Proceeding as in Ref. \cite{Fisher1994}, we derive the equation for this moment by multiplying Eq. \eqref{eq:general-flow} by $\b$ and then integrating over $\b$
\be\label{eq:first-moment-flow}
%\begin{split}
\partial_\Gamma \ln\left({\bar\b}(\zeta)\,\rho \right) =
\partial_\zeta \ln\left({\bar\b}(\zeta)\, \rho\right)
+ \frac{2 a(\Gamma)}{{\bar\b}(\zeta)} \int_0^\zeta d\zeta' {\bar\b}(\zeta')
%\end{split}
\ee
Note that we have neglected the $\ln 4$ in \eqref{eq:general-flow}. This is justified near the fixed point since the typical $\b$ flows to $\infty$. After substituting the solution for $\rho(\zeta, \Gamma)$ in Eq. \eqref{eq:first-moment-flow} we find the solution
\be\label{solution2}
{\bar\b}(\zeta)= \frac{1}{b_0} (a_0 \Gamma +1)^\phi \left(1+\frac{\zeta \phi}{\Gamma+\frac{1}{a_0}}\right)
\ee
where $\phi=(1+\sqrt{5})/2$ is the golden ratio and $b_0$ is determined by the initial condition.

\begin{figure}[t]
 \centerline{\resizebox{0.9\linewidth}{!}{\includegraphics{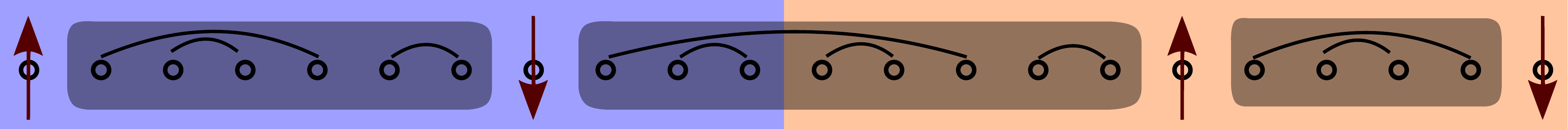}}}
 \caption{{\em Schematic illustration of remaining spins and clusters of decimated pairs in the renormalized chain at time $t$.} }
 \label{fig:clusters}
\end{figure}

Another important ingredient for calculation of physical properties is the distance between remaining spins (i.e. length of decimated clusters) at time $t$. An illustration of such clusters is shown in Fig. \ref{fig:clusters}.
Since the RG flow is the same as in Ref. [\onlinecite{Fisher1994}] we similarly obtain
$L_\Gamma = (a_0 \Gamma+1)^2 = \left[a_0\ln(\Omega_0 t) + 1 \right]^2$
Which behaves as $\ln^2 (\W_0 t)$ at long times. From this result we can immediately infer the decay of the staggered moment which is simply the fraction of remaining spins at time $t$: $m_s=1/L(t)=1/\left(a_0\ln(\Omega_0 t) + 1 \right)^2$ . It is interesting to contrast this behavior with the decay of the staggered moment in the analogous quench of a clean $XXZ$ model. In that case Ref. [\onlinecite{Barmettler2010}] found oscillatory decay of the staggered moment (for $\D<1$) with an envelope that decays exponentially in time.

We note that in the {\em non-interacting} case $\D_i=0$ the RG flow generalizes in a straight forward way to an arbitrary Ising initial states rather than a Ne\'el state.
If the strongest exchange coupling happens to be between spins with the same orientations, this pair will freeze in their state, while quantum fluctuations lead to an effective coupling between the neighbors that is the same in absolute value as that obtained for anti-aligned spins. Hence the distribution of the absolute values of the exchange couplings is governed by the same RG equations. In the interacting case, such aligned pairs lead to further complications that are left for later work.

%Another important consequence of the RG time evolution is the presence of integrals of motion. Each engaged pair keeps oscillating between the two oppositely aligned pair configurations. Hence, the operator $S_1^z S_2^z$, which is the product of the z-components of the spins which are strongly coupled, is a constant of motion, with the eigenvalue $-1/4$. At the first stages of the RG flow, where the perturbation theory is not yet fully controlled, these quantities are an approximate integrals of motion for short times. However, when the RG flow approaches the fixed point distribution, and the spin pairs become more distant according to Eq. \eqref{eq:L}, the integrals of motion become gradually better, and essentially asymptotically exact.
%\section{Evolution of entanglement entropy}\label{sec:entanglement-entropy}
\noindent
{\em Results --} We now use the solutions of the flow equations to compute the time evolution of important physical quantities. Specifically to gain information on particle transport and on thermalization of the system we consider the growth of the total particle number fluctuation and of the entanglement entropy in a sub-system corresponding to half of the chain.

To compute the particle number fluctuations we note that the total particle number (or $S^z_{\text{tot}}$) within a decimated pair of sites is a conserved quantity in the RG scheme. Therefore the only contribution to the particle number fluctuation in one half of the system comes from decimated pairs that reside on different sides of the interface. Each oscillating pair that cuts the interface adds $1/8$ to the number fluctuation on time average. Computing the total particle number fluctuation then amounts to counting the number of decimated bonds that cuts the interface. Proceeding exactly as in the random singlet phase\cite{Refael2004} we have
$N_p \approx \int^\G d\Gamma' a(\Gamma') = \ln\left(\Gamma+1/a_0\right)$. Hence the particle number fluctuation grows extremely slowly as $\av{\d N^2}= (1/24)\ln(\ln(\W_0 t))$ at long times. Interestingly this result is independent of the interaction strength $\D$.

We now turn to the growth of the entanglement entropy,between the two halves ($A$ and $B$) of the system, $S\equiv -\tr \rho_A\log_2\rho_A$. Consider first the simpler (non-interacting) case
$\D_i=0$, where no entanglement is generated between a decimated pair and the rest of the system. Then, the only source of entanglement between the two halves of the system are decimated pairs that cut the interface, each contributes a time average of $S_p=2-1/\ln 2\approx 0.557$. The growth of the entropy is therefore similar to that of the particle number fluctuation
\be \label{eq:S_0-evolution}
S_0(t) \approx S_p\frac{1}{3} \ln\left(\ln\left(\Omega_0 t\right)+1/a_0\right)
\ee
We can generalize this result (for $\D_i=0$) to a quench from an arbitrary Ising state with a fraction $q$ of anti-aligned neighbors. Because $q$ is an invariant of the RG and aligned pairs do not contribute to the entropy the prefactor in (\ref{eq:S_0-evolution}) changes to $q S_p$.

We turn to our main focus, which is the interacting system with $\D \neq 0$. In this case a pair decimated at time $t_1$ will eventually get entangled with the neighboring spins according to Eq. \eqref{eq:Heff} after a characteristic time  $t_\text{ent}(t_1)=2\W_1/(J_1^2 \D_1)$. In particular from the time of the quench, entanglement will be generated by the interaction only after a delay time $t_{\text{delay}}=2\W_0/(J_0^2\D_0)= 2(\W_0/J_0)(1/J^z_0)$, where $J^z_0\equiv J_0\D_0$ is the typical value of the bare interaction energy.

The interaction-generated entanglement entropy found at time $t$ originates from entanglement of pairs eliminated at an earlier time $t_1=t-t_{\text{ent}}$, which corresponds to the scaling parameter $\G_1=\ln \W_0 t_1$. To estimate this contribution to the entropy we recall that spins on the renormalized chain at time $t_1$ are separated by clusters of length $L(\G_1)$ of spins decimated at even earlier times and are oscillating at higher frequencies. By the time $t$ that a pair of spins decimated at $t_1$ entangles with their neighbors, the spins inside the decimated clusters must also be entangled with each other. So, it is safe to assume that by the observation time $t$ entanglement propagates to a distance $L(\G_1)$ giving rise to entanglement entropy $S\approx 0.5 L(\G_1)\approx 0.5 \,(a_0 \G_1+1)^2$. The factor $0.5$ stems from the number of available degrees of freedom: the two states with aligned spins in each decimated pair remain unpopulated and therefore do not contribute to the entropy. To write this as a function of the time $t$ we use the relation between $t$ and $t_1$,
\be
t = t_1 + t_{\text{ent}}  =  t_1\left(1+ \frac{2\Omega_1^2}{J_L  J_R \D_1}\right)
\approx t_1 \frac{2\Omega_1^2}{J_L J_R\D_1}.
\ee
We now take the logarithm of both sides and replace the scaling variables by their appropriate average values $\zeta\to 1/a(\G_1) $ and $\b_1\to {\bar\b}(\zeta=0;\G_1)$. Here the correlations between the random variables came in to play. We needed the average of $\b$ on the bonds with strongest $J$ ($\zeta=0$), which is different than the global average of $\b$.

Using the solutions (\ref{eq:solution}) and \eqref{solution2} for the typical values we find
$
\G=3 \G_1 +\frac{1}{b_0} (a_0 \G_1 +1)^\phi + 2/a_0 + \ln2,
$
This equation should be inverted to express $\G_1$ as a function of $\G$ and used to obtain  $S(\G)=0.5 L(\G_1(\G))$.
In general, the inversion cannot be done analytically, but we can easily find the limiting regimes.  At sufficiently long times the term $\G_1^\phi$ dominates the right hand side and then we have $a_0 \G_1= [b_0 (\G-2/a_0-\ln2)]^{1/\phi} -1$. On the other hand at shorter times, when the linear term dominates,  we get $\G_1 = {1\over 3}(\G - {2\over a_0} - {1\over b_0} - \ln2)$.

The crossover time $t^*$ separating the two regimes depends on the initial conditions through the coefficients of the terms $\G_1$ and $\G_1^\phi$.
If $b_0\gg a_0$, that is for stronger disorder in hopping than in the interactions, we have  $t^* =  t_{\text{delay}} \exp \left(6(3 b_0/a_0)^\phi /a_0\right)$. In the opposite regime $b_0\ll a_0$ the term $\G_1^\phi$  dominates from the outset and $t^* = t_{\text{delay}}$. We can now write an expression for the
 growth of the entanglement entropy valid in the limiting regimes
\be
\begin{split}
S(t)&\approx {1\over 2} \left(\frac{\ln\left(t/t_{\text{delay}} \right)}{\ln \left(\W_0/J_0\right)}+1\right)^2\t(t-t_{\text{delay}})\t(t^*-t) \\
&\quad + {1\over 2} \left(\frac{\ln\left(t/t_{\text{delay}} \right)}{\ln \left(1/\D_0\right)}+1\right)^{2/\phi}\t(t-t^*)-{1\over 2}.
\end{split}
\label{eq:S}
\ee
Interestingly, Eq. (\ref{eq:S}) gives the unbounded logarithmic growth of the entanglement entropy seen in the numerical simulations and even the delay of this interaction induced growth by a time that scales as the inverse interaction strength [\onlinecite{Bardarson2012}]. It is important to note however that the numerically simulated model differs from ours by having disordered Zeeman coupling, which may ultimately drive it to a different fixed point \cite{Ronen-future}. Nevertheless we can attempt a comparison even of the more subtle form of the logarithmic growth by crudely accounting for this difference. Through second order perturbation theory the Zeeman strong disorder generates predominantly strong randomness in the hopping, but not in the interactions. Hence the natural regime to compare the numerics to our model is $a_0\ll b_0$ where the first line of (\ref{eq:S}) holds at any reasonable time-scales (i.e. below the exponentially long time $t^*$). Even within this regime Eq. (\ref{eq:S}) predicts a crossover between growth of the entanglement entropy as $\ln t$ at times $t<(\W_0/J_0)^2 t_{\text{delay}}$ to growth as $\ln^2 t$ at longer times. The slope of the logarithmic growth $1/\ln(\W_0/J_0)$ is completely independent of the interaction strength $\D$, consistent with the universality
identified in Ref. [\onlinecite{Bardarson2012}].
%In second order perturbation theory the random Zeeman fields included in [\onlinecite{Bardarson2012}] generate strong disorder in the hopping, but not in the interaction. Therefore it is reasonable to compare these calculations with our regime $a_0\ll b_0$.

To complete the comparison, we note that the growth of particle number fluctuations which was computed above is much slower  ($\sim \ln\ln t$) than the growth of the entanglement entropy. Again this is consistent with the numerical results, which implies that particle transport is essentially blocked. Indeed the particle number fluctuations are in general only a lower bound of the entanglement entropy\cite{Klich2006}.

Having found that the entanglement entropy increases without bound it is natural to ask if this leads to thermalization. To address this issue let us consider the saturation of the entanglement entropy in a finite system, or in a finite sub-system of length $L_s$.  Eq. (\ref{eq:S}) implies that the entropy will approach its maximal value $S_\infty$ after a time $t_{\text{sat}}\approx t_{delay}\exp\left[-\ln\left(\D_0\right)L_s^{\phi/2}\right]$.
Does the saturation value $S_\infty$ correspond to a state in thermal equilibrium?

Provided we start from a symmetric distribution of $\D_i$ such that $\av{\D_i}=0$, then the initial Ne\'el state has zero mean energy, exactly in the middle of the many-body energy spectrum. If this state thermalized following the quench, the entanglement entropy would have to saturate to its infinite temperature value of $L$. But as we have pointed out above, the RG flow implies a saturation entropy that is at most half of the infinite temperature value because half of the degrees of freedom remain frozen in the dynamics. This fact is embodied in the infinite set of emergent  integrals of motion $I_p=(S^z_1 S^z_2)_p$, which account for the fact that a pair of decimated spins, never flip their relative orientation within the perturbative RG scheme. Note that this remains true even for generalized initial states allowing for aligned neighboring spins. In the case of an initial Ne\'el state we also have the particle number on oscillating pairs (or $(S^z_1+S^z_2)_p$) as additional integrals of motion. These emergent conservation laws become asymptotically exact for well separated pairs of spins decimated at long times as the perturbative RG scheme becomes asymptotically exact near the infinite randomness fixed point.
We conclude that the long time steady state of the chain with non vanishing interaction is characterized by the generalized Gibbs ensemble (GGE)\cite{Rigol2008}, which describes thermalization within a subspace constrained by the values of the emergent integrals of motion $I_p$.

It is interesting to note that the long time steady state attained by the non interacting state $\D_i=0$ is markedly different. The extremely slow increase of the entanglement entropy as $\ln\ln t$ given by Eq. (\ref{eq:S_0-evolution}) together with the relation between length and time scales $\ln \W_0 t=\G\sim \sqrt{L}$ imply saturation of the entanglement entropy to $S_\infty\approx {S_p\over 6}\ln L$. This result, as well as the $\ln\ln t$ growth of the entropy matches with numerical results obtained for the random transverse field Ising chain\cite{Igloi2012} that can be similarly described by a model of noninteracting fermions.

So far we have used the RG approach to establish and characterize a many-body localized state. It is interesting to examine the criterion for validity of the RG scheme, which may indicate a transition to a different, possibly thermalizing and delocalized, state. Such a criterion can be obtained from the first term in Eq. (\ref{eq:Heff}). The perturbation theory at the base of the renormalization step is valid while $(J_L J_R/\W)\ll \av{J} (1-\D_S^2)$. This can be expressed in terms of the average values of parameters at the beginning of the flow. Using $\av{J_0/\W_0}=a_0/(a_0+1)$, we obtain the criterion
$
a_0< \D_0^{-2}-1
$.
Recall that increasing $a_0$ corresponds to decreasing disorder. It is tempting to speculate that for weaker disorder $a_0>\D^{-2}-1$ the dynamics will give rise to a normal thermalizing state. However the above requirement may be too strict. Even if the criterion $a_0<\D_0^{-2}-1$ is not satisfied initially, this may be corrected in later stages of the flow as $\D$ independently flows to smaller values. It is therefore also possible that the transition to a different state occurs only for $\av{\D^2}\approx 1$ independent of the disorder strength.

%In Fig. \ref{fig:staggered} we plot the phase diagram based on this criterion with a line marking the conjectured location of the many body localization transition. The y-axis in this figure represents the inverse of the disorder strength whereas the x-axis is the typical interaction strength. Note that this phase diagram pertains to the class of models discussed here, namely a random xxz model, with no Zeeman fields, which is quenched from a a staggered (antiferromagnetic) initial state.

%We note that for an arbitrary initial Ising state with $\D=0$ the RG can be carried out in the same way. In this case, if the strongest exchange coupling is between spins with the same orientations, these spins will be frozen. Quantum fluctuations generate an effective coupling between the neighbors similar to the one obtained in the opposite orientation, with a minus sign. Therefore, the distribution of the absolute value of the exchange couplings is governed by the same RG equations.

{\em Conclusion --} Using a real space RG scheme formulated in real time, we gave a dynamical description of a many-body localized state in a random spin chain, equivalent to interacting fermions with random hopping. Within this approach the localized state is characterized by a flow to an infinite randomness fixed point. Solution of the flow equations allows us to characterize this state in a rather detailed way. The results are found to be in excellent agreement with recent numerical simulations done on a similar, albeit not identical, model\cite{Bardarson2012}.

Particle localization is manifest in the extremely slow growth $\sim \ln\ln t$ of the particle number fluctuations in half the system that is seen both in the interacting and non-interacting systems. The entanglement entropy $S$ reveals a dramatic difference between the Anderson localized state of noninteracting fermions and the many-body localized state established when interactions are present. In the non interacting system $S$ grows together with the particle number fluctuation as $S(t)\sim \log\log t$ and saturates to a non-extensive value $\sim \ln L$ in a finite system. Interactions lead to much faster growth of $S(t)$ as $\log^{2/\phi} t$ at long times, but they take effect only after a delay time that scales as the inverse of the interaction strength $t_{\text{delay}}\sim 1/J^z$. Furthermore, the $\log^{2/\phi} t$ behavior seen in the long time limit is preceded by $\log t$ growth up to an intermediate time scale $t_{\text{lin}}\sim t_{\text{delay}} (\W_0/J_0)^2\gg t_{\text{delay}}$.
It is interesting to note that the growth of entanglement as $\ln^{2/\phi} t$ exceeds the upper bound $\sim \ln t$ proved for non-interacting Anderson localized chains\cite{Burrell2007}.

The RG flow toward the infinite randomness fixed point has direct consequences on the equilibration in this system. In a sub-system of length $L$ the entanglement entropy saturates to an extensive value $S_\infty \sim L$, which is however smaller than it would reach had the system attained true thermal equilibrium.
We attribute the lack of thermalization to an infinite set of emergent integrals of motion, which become asymptotically exact conservation laws near the infinite randomness dynamical fixed point. The dynamics of the system can therefore be viewed as
thermalization within a GGE characterized by the emergent set of conserved quantities, a possibility suggested in Ref. [\onlinecite{Gogolin2011},\onlinecite{{Khatami2012}}] . Here we demonstrated that such a GGE emerges in a non integrable random system as a dynamical fixed point of the renormalization group and captures the essence of a many-body localized state.
The nature of the critical point marking the transition to the normal thermalizing state remains an interesting question for future study as are generalizations of our scheme to more generic disorder models and initial states.

{Acknowledgements --} We thank J. E. Moore, F. Pollmann, D. Huse, V. Oganesyan, G. Refael, and A. Polkovnikov for useful discussions. This work was supported by the ISF and the Minerva foundation.

\bibliographystyle{phd-url-notitle}
\bibliography{mbl}

\begin{thebibliography}{24}
\expandafter\ifx\csname natexlab\endcsname\relax\def\natexlab#1{#1}\fi
\expandafter\ifx\csname bibnamefont\endcsname\relax
  \def\bibnamefont#1{#1}\fi
\expandafter\ifx\csname bibfnamefont\endcsname\relax
  \def\bibfnamefont#1{#1}\fi
\expandafter\ifx\csname citenamefont\endcsname\relax
  \def\citenamefont#1{#1}\fi
\expandafter\ifx\csname url\endcsname\relax
  \def\url#1{\texttt{#1}}\fi
\expandafter\ifx\csname urlprefix\endcsname\relax\def\urlprefix{URL }\fi
\providecommand{\bibinfo}[2]{#2}
\providecommand{\eprint}[2][]{\url{#2}}

\bibitem[{\citenamefont{Anderson}(1958)}]{Anderson1958}
\bibinfo{author}{\bibfnamefont{P.~W.} \bibnamefont{Anderson}},
  \bibinfo{journal}{Phys. Rev.} \textbf{\bibinfo{volume}{109}},
  \bibinfo{pages}{1492} (\bibinfo{year}{1958}),
  \href{http://link.aps.org/doi/10.1103/PhysRev.109.1492}{URL}.

\bibitem[{\citenamefont{Basko et~al.}(2006)\citenamefont{Basko, Aleiner, and
  Altshuler}}]{Basko2006}
\bibinfo{author}{\bibfnamefont{D.}~\bibnamefont{Basko}},
  \bibinfo{author}{\bibfnamefont{I.}~\bibnamefont{Aleiner}}, \bibnamefont{and}
  \bibinfo{author}{\bibfnamefont{B.}~\bibnamefont{Altshuler}},
  \bibinfo{journal}{Annals of Physics} \textbf{\bibinfo{volume}{321}},
  \bibinfo{pages}{1126} (\bibinfo{year}{2006}), ISSN \bibinfo{issn}{00034916},
  \href{http://linkinghub.elsevier.com/retrieve/pii/S0003491605002630}{URL}.

\bibitem[{\citenamefont{Oganesyan and Huse}(2007)}]{Oganesyan2007}
\bibinfo{author}{\bibfnamefont{V.}~\bibnamefont{Oganesyan}} \bibnamefont{and}
  \bibinfo{author}{\bibfnamefont{D.}~\bibnamefont{Huse}},
  \bibinfo{journal}{Physical Review B} \textbf{\bibinfo{volume}{75}},
  \bibinfo{pages}{1} (\bibinfo{year}{2007}), ISSN \bibinfo{issn}{1098-0121},
  \href{http://link.aps.org/doi/10.1103/PhysRevB.75.155111}{URL}.

\bibitem[{\citenamefont{Monthus and Garel}(2010)}]{Monthus2010}
\bibinfo{author}{\bibfnamefont{C.}~\bibnamefont{Monthus}} \bibnamefont{and}
  \bibinfo{author}{\bibfnamefont{T.}~\bibnamefont{Garel}},
  \bibinfo{journal}{Physical Review B} \textbf{\bibinfo{volume}{81}},
  \bibinfo{pages}{134202} (\bibinfo{year}{2010}), \eprint{arXiv:1001.2984v2},
  \href{http://prb.aps.org/abstract/PRB/v81/i13/e134202}{URL}.

\bibitem[{\citenamefont{Pal and Huse}(2010)}]{Pal2010}
\bibinfo{author}{\bibfnamefont{A.}~\bibnamefont{Pal}} \bibnamefont{and}
  \bibinfo{author}{\bibfnamefont{D.}~\bibnamefont{Huse}},
  \bibinfo{journal}{Physical Review B} \textbf{\bibinfo{volume}{82}},
  \bibinfo{pages}{1} (\bibinfo{year}{2010}), ISSN \bibinfo{issn}{1098-0121},
  \href{http://link.aps.org/doi/10.1103/PhysRevB.82.174411}{URL}.

\bibitem[{\citenamefont{Canovi et~al.}(2011)\citenamefont{Canovi, Rossini,
  Fazio, Santoro, and Silva}}]{Canovi2011}
\bibinfo{author}{\bibfnamefont{E.}~\bibnamefont{Canovi}},
  \bibinfo{author}{\bibfnamefont{D.}~\bibnamefont{Rossini}},
  \bibinfo{author}{\bibfnamefont{R.}~\bibnamefont{Fazio}},
  \bibinfo{author}{\bibfnamefont{G.}~\bibnamefont{Santoro}}, \bibnamefont{and}
  \bibinfo{author}{\bibfnamefont{A.}~\bibnamefont{Silva}},
  \bibinfo{journal}{Physical Review B} \textbf{\bibinfo{volume}{83}},
  \bibinfo{pages}{1} (\bibinfo{year}{2011}), ISSN \bibinfo{issn}{1098-0121},
  \href{http://link.aps.org/doi/10.1103/PhysRevB.83.094431}{URL}.

\bibitem[{\citenamefont{Chiara et~al.}(2006)\citenamefont{Chiara, Montangero,
  Calabrese, and Fazio}}]{Chiara2006}
\bibinfo{author}{\bibfnamefont{G.~D.} \bibnamefont{Chiara}},
  \bibinfo{author}{\bibfnamefont{S.}~\bibnamefont{Montangero}},
  \bibinfo{author}{\bibfnamefont{P.}~\bibnamefont{Calabrese}},
  \bibnamefont{and} \bibinfo{author}{\bibfnamefont{R.}~\bibnamefont{Fazio}},
  \bibinfo{journal}{Journal of Statistical Mechanics: Theory and Experiment}
  \textbf{\bibinfo{volume}{2006}}, \bibinfo{pages}{P03001}
  (\bibinfo{year}{2006}), ISSN \bibinfo{issn}{1742-5468},
  \href{http://stacks.iop.org/1742-5468/2006/i=03/a=P03001?key=crossref.304a80d96bf6f2f5edaeff533c20f096}{URL}.

\bibitem[{\citenamefont{\v{Z}nidari\v{c}
  et~al.}(2008)\citenamefont{\v{Z}nidari\v{c}, Prosen, and
  Prelov\v{s}ek}}]{Znidaric2008}
\bibinfo{author}{\bibfnamefont{M.}~\bibnamefont{\v{Z}nidari\v{c}}},
  \bibinfo{author}{\bibfnamefont{T.}~\bibnamefont{Prosen}}, \bibnamefont{and}
  \bibinfo{author}{\bibfnamefont{P.}~\bibnamefont{Prelov\v{s}ek}},
  \bibinfo{journal}{Physical Review B} \textbf{\bibinfo{volume}{77}},
  \bibinfo{pages}{1} (\bibinfo{year}{2008}), ISSN \bibinfo{issn}{1098-0121},
  \href{http://link.aps.org/doi/10.1103/PhysRevB.77.064426}{URL}.

\bibitem[{\citenamefont{{Bardarson} et~al.}(2012)\citenamefont{{Bardarson},
  {Pollmann}, and {Moore}}}]{Bardarson2012}
\bibinfo{author}{\bibfnamefont{J.~H.} \bibnamefont{{Bardarson}}},
  \bibinfo{author}{\bibfnamefont{F.}~\bibnamefont{{Pollmann}}},
  \bibnamefont{and} \bibinfo{author}{\bibfnamefont{J.~E.}
  \bibnamefont{{Moore}}}, \bibinfo{journal}{ArXiv e-prints}
  (\bibinfo{year}{2012}), \eprint{1202.5532}.

\bibitem[{\citenamefont{Dasgupta and Ma}(1980)}]{Dasgupta1980}
\bibinfo{author}{\bibfnamefont{C.}~\bibnamefont{Dasgupta}} \bibnamefont{and}
  \bibinfo{author}{\bibfnamefont{S.-k.} \bibnamefont{Ma}},
  \bibinfo{journal}{Physical Review B} \textbf{\bibinfo{volume}{22}},
  \bibinfo{pages}{1305} (\bibinfo{year}{1980}), ISSN \bibinfo{issn}{0163-1829},
  \href{http://prb.aps.org/abstract/PRB/v22/i3/p1305\_1
  http://link.aps.org/doi/10.1103/PhysRevB.22.1305}{URL}.

\bibitem[{\citenamefont{Bhatt and Lee}(1982)}]{Bhatt1982}
\bibinfo{author}{\bibfnamefont{R.}~\bibnamefont{Bhatt}} \bibnamefont{and}
  \bibinfo{author}{\bibfnamefont{P.}~\bibnamefont{Lee}},
  \bibinfo{journal}{Physical Review Letters} \textbf{\bibinfo{volume}{48}},
  \bibinfo{pages}{344} (\bibinfo{year}{1982}), ISSN \bibinfo{issn}{0031-9007},
  \href{http://prl.aps.org/abstract/PRL/v48/i5/p344\_1}{URL}.

\bibitem[{\citenamefont{Fisher}(1994)}]{Fisher1994}
\bibinfo{author}{\bibfnamefont{D.}~\bibnamefont{Fisher}},
  \bibinfo{journal}{Physical Review B} \textbf{\bibinfo{volume}{50}},
  \bibinfo{pages}{3799} (\bibinfo{year}{1994}),
  \href{http://prb.aps.org/abstract/PRB/v50/i6/p3799\_1}{URL}.

\bibitem[{\citenamefont{Fisher et~al.}(1998)\citenamefont{Fisher, {Le Doussal},
  and Monthus}}]{Fisher1998}
\bibinfo{author}{\bibfnamefont{D.}~\bibnamefont{Fisher}},
  \bibinfo{author}{\bibfnamefont{P.}~\bibnamefont{{Le Doussal}}},
  \bibnamefont{and} \bibinfo{author}{\bibfnamefont{C.}~\bibnamefont{Monthus}},
  \bibinfo{journal}{Physical Review Letters} \textbf{\bibinfo{volume}{80}},
  \bibinfo{pages}{3539} (\bibinfo{year}{1998}), ISSN \bibinfo{issn}{0031-9007},
  \href{http://link.aps.org/doi/10.1103/PhysRevLett.80.3539}{URL}.

\bibitem[{\citenamefont{Lee et~al.}(2009)\citenamefont{Lee, Refael, Cross,
  Kogan, and Rogers}}]{Lee2009}
\bibinfo{author}{\bibfnamefont{T.~E.} \bibnamefont{Lee}},
  \bibinfo{author}{\bibfnamefont{G.}~\bibnamefont{Refael}},
  \bibinfo{author}{\bibfnamefont{M.~C.} \bibnamefont{Cross}},
  \bibinfo{author}{\bibfnamefont{O.}~\bibnamefont{Kogan}}, \bibnamefont{and}
  \bibinfo{author}{\bibfnamefont{J.~L.} \bibnamefont{Rogers}},
  \bibinfo{journal}{Phys. Rev. E} \textbf{\bibinfo{volume}{80}},
  \bibinfo{pages}{046210} (\bibinfo{year}{2009}),
  \href{http://link.aps.org/doi/10.1103/PhysRevE.80.046210}{URL}.

\bibitem[{\citenamefont{Mathey and Polkovnikov}(2010)}]{Mathey2010}
\bibinfo{author}{\bibfnamefont{L.}~\bibnamefont{Mathey}} \bibnamefont{and}
  \bibinfo{author}{\bibfnamefont{a.}~\bibnamefont{Polkovnikov}},
  \bibinfo{journal}{Physical Review A} \textbf{\bibinfo{volume}{81}},
  \bibinfo{pages}{1} (\bibinfo{year}{2010}), ISSN \bibinfo{issn}{1050-2947},
  \href{http://link.aps.org/doi/10.1103/PhysRevA.81.033605}{URL}.

\bibitem[{\citenamefont{Barmettler et~al.}(2010)\citenamefont{Barmettler, Punk,
  Gritsev, Demler, and Altman}}]{Barmettler2010}
\bibinfo{author}{\bibfnamefont{P.}~\bibnamefont{Barmettler}},
  \bibinfo{author}{\bibfnamefont{M.}~\bibnamefont{Punk}},
  \bibinfo{author}{\bibfnamefont{V.}~\bibnamefont{Gritsev}},
  \bibinfo{author}{\bibfnamefont{E.}~\bibnamefont{Demler}}, \bibnamefont{and}
  \bibinfo{author}{\bibfnamefont{E.}~\bibnamefont{Altman}},
  \bibinfo{journal}{New Journal of Physics} \textbf{\bibinfo{volume}{12}},
  \bibinfo{pages}{055017} (\bibinfo{year}{2010}), ISSN
  \bibinfo{issn}{1367-2630},
  \href{http://stacks.iop.org/1367-2630/12/i=5/a=055017?key=crossref.3018bc54b522a24e6c3414e81e260a2b}{URL}.

\bibitem[{\citenamefont{Refael and Moore}(2004)}]{Refael2004}
\bibinfo{author}{\bibfnamefont{G.}~\bibnamefont{Refael}} \bibnamefont{and}
  \bibinfo{author}{\bibfnamefont{J.}~\bibnamefont{Moore}},
  \bibinfo{journal}{Physical Review Letters} \textbf{\bibinfo{volume}{93}},
  \bibinfo{pages}{1} (\bibinfo{year}{2004}), ISSN \bibinfo{issn}{0031-9007},
  \href{http://link.aps.org/doi/10.1103/PhysRevLett.93.260602}{URL}.

\bibitem[{\citenamefont{Vosk and Altman}()}]{Ronen-future}
\bibinfo{author}{\bibfnamefont{R.}~\bibnamefont{Vosk}} \bibnamefont{and}
  \bibinfo{author}{\bibfnamefont{E.}~\bibnamefont{Altman}}, \bibinfo{note}{to
  be published}.

\bibitem[{\citenamefont{Klich et~al.}(2006)\citenamefont{Klich, Refael, and
  Silva}}]{Klich2006}
\bibinfo{author}{\bibfnamefont{I.}~\bibnamefont{Klich}},
  \bibinfo{author}{\bibfnamefont{G.}~\bibnamefont{Refael}}, \bibnamefont{and}
  \bibinfo{author}{\bibfnamefont{A.}~\bibnamefont{Silva}},
  \bibinfo{journal}{Phys. Rev. A} \textbf{\bibinfo{volume}{74}},
  \bibinfo{pages}{032306} (\bibinfo{year}{2006}),
  \href{http://link.aps.org/doi/10.1103/PhysRevA.74.032306}{URL}.

\bibitem[{\citenamefont{Rigol et~al.}(2008)\citenamefont{Rigol, Dunjko, and
  Olshanii}}]{Rigol2008}
\bibinfo{author}{\bibfnamefont{M.}~\bibnamefont{Rigol}},
  \bibinfo{author}{\bibfnamefont{V.}~\bibnamefont{Dunjko}}, \bibnamefont{and}
  \bibinfo{author}{\bibfnamefont{M.}~\bibnamefont{Olshanii}},
  \bibinfo{journal}{Nature} \textbf{\bibinfo{volume}{452}},
  \bibinfo{pages}{854} (\bibinfo{year}{2008}), ISSN \bibinfo{issn}{1476-4687},
  \href{http://www.ncbi.nlm.nih.gov/pubmed/18421349}{URL}.

\bibitem[{\citenamefont{Igl\'{o}i et~al.}(2012)\citenamefont{Igl\'{o}i,
  Szatm\'{a}ri, and Lin}}]{Igloi2012}
\bibinfo{author}{\bibfnamefont{F.}~\bibnamefont{Igl\'{o}i}},
  \bibinfo{author}{\bibfnamefont{Z.}~\bibnamefont{Szatm\'{a}ri}},
  \bibnamefont{and} \bibinfo{author}{\bibfnamefont{Y.-C.} \bibnamefont{Lin}},
  \bibinfo{journal}{Physical Review B} \textbf{\bibinfo{volume}{85}}
  (\bibinfo{year}{2012}), ISSN \bibinfo{issn}{1098-0121},
  \href{http://prb.aps.org/abstract/PRB/v85/i9/e094417}{URL}.

\bibitem[{\citenamefont{Burrell and Osborne}(2007)}]{Burrell2007}
\bibinfo{author}{\bibfnamefont{C.}~\bibnamefont{Burrell}} \bibnamefont{and}
  \bibinfo{author}{\bibfnamefont{T.}~\bibnamefont{Osborne}},
  \bibinfo{journal}{Physical Review Letters} \textbf{\bibinfo{volume}{99}},
  \bibinfo{pages}{1} (\bibinfo{year}{2007}), ISSN \bibinfo{issn}{0031-9007},
  \href{http://link.aps.org/doi/10.1103/PhysRevLett.99.167201}{URL}.

\bibitem[{\citenamefont{Gogolin et~al.}(2011)\citenamefont{Gogolin, M\"{u}ller,
  and Eisert}}]{Gogolin2011}
\bibinfo{author}{\bibfnamefont{C.}~\bibnamefont{Gogolin}},
  \bibinfo{author}{\bibfnamefont{M.}~\bibnamefont{M\"{u}ller}},
  \bibnamefont{and} \bibinfo{author}{\bibfnamefont{J.}~\bibnamefont{Eisert}},
  \bibinfo{journal}{Physical Review Letters} \textbf{\bibinfo{volume}{106}},
  \bibinfo{pages}{1} (\bibinfo{year}{2011}), ISSN \bibinfo{issn}{0031-9007},
  \href{http://link.aps.org/doi/10.1103/PhysRevLett.106.040401}{URL}.

\bibitem[{\citenamefont{Khatami et~al.}(2012)\citenamefont{Khatami, Rigol,
  Relano, and Garcia-Garcia}}]{Khatami2012}
\bibinfo{author}{\bibfnamefont{E.}~\bibnamefont{Khatami}},
  \bibinfo{author}{\bibfnamefont{M.}~\bibnamefont{Rigol}},
  \bibinfo{author}{\bibfnamefont{A.}~\bibnamefont{Relano}}, \bibnamefont{and}
  \bibinfo{author}{\bibfnamefont{A.~M.} \bibnamefont{Garcia-Garcia}},
  \bibinfo{journal}{Phys. Rev. E} \textbf{\bibinfo{volume}{85}},
  \bibinfo{pages}{050102} (\bibinfo{year}{2012}),
  \href{http://link.aps.org/doi/10.1103/PhysRevE.85.050102}{URL}.

\end{thebibliography}

\end{document}